\documentclass[12pt]{article}

\usepackage{axodraw}
\newcommand{\GeV}{\,\mbox{GeV}}
\newcommand{\dx}{\mathrm{dx}\,}
\newcommand{\dy}{\mathrm{dy}\,}
\newcommand{\nnb}{\nonumber}

\newcommand{\al}{\alpha} 
\newcommand{\ice}[1]{\relax}
\newcommand{\be}{\begin{equation}}
\newcommand{\ee}{\end{equation}}

\newcommand{\g}{\gamma}
\newcommand{\msbar}{\overline{\mbox{MS}}}
\begin{document}
\thispagestyle{empty}
\begin{flushright}
\centerline{\normalsize\hfill hep-ph/0105093}
\end{flushright}
\vspace*{5mm}
\begin{center}
{\Large\bf Vacuum Saturation Hypothesis and QCD Sum Rules\footnote{\normalsize
Copy of the
paper  published in {\it IL NUOVO CIMENTO}, {\bf 100 A} (1988) 899.
Received 28 March 1988. Some comments, related to a (relatively
small) change of the updated result due to changes in the values 
of incorporated phenomenological parameters, are added.}}
\\
\vspace*{1cm}
{\bf K.G. Chetyrkin and A.A. Pivovarov}\\
\vspace{0.3cm}
Institute for Nuclear Research of the Academy of Sciences of Russian 
Federation,\\
60th October Anniversary prospect, 7a, 117312 Moscow, Russian Federation
\end{center}

\begin{abstract}
The accuracy of the vacuum saturation hypothesis is discussed
using the examples of vacuum expectation values of four-quark
operators
and the parameter $B$, which determines the short-distance
contribution to the $K^0 - \bar K^0$ mixing.                                    
\end{abstract}


\newpage 
1. - The standard model of strong and electroweak interactions
predicts the existence of transitions changing the strangeness by two
units. These processes are due to exchange of intermediate vector
bosons and lead to a possibility of $K^0 - \bar K^0$ mixing. A
calculation of the corresponding effective Hamiltonian on the basis of
the standard model was first performed by Gaillard and Lee \cite{1} by
expanding the interaction in the inverse mass of the intermediate
boson. The comparison of the result of this calculation with the
experimentally observed mass difference of $K_L$ and $K_S$ mesons led
to a prediction for the c-quark mass before experimental detection of
$J/\psi$ particle. In spite of the considerable theoretical
uncertainty, involved in evaluation of the matrix element $\langle
K^0| H_{eff}|\bar K^0 \rangle$ the predicted quark mass happened to be
in rather good agreement with the measured one.  However, the value of
the $K_L-K_S$ mass difference is quite sensitive to the numerical
values of such parameters of the standard model as the masses of the
t- and c-quarks, the mixing angles between various quark generations
and the general structure of the model. Since the pioneer
calculation by Gaillard and Lee, a good deal of work has been
carried out to refine its results.

Account was taken of the strong-interaction corrections to $H_{eff}$
in the  leading
logs approximation and for the three-quark generations by
Gillman and Wise \cite{2}. The obtained expression for the
$K_L-K_S$ mass difference is of the
form
\[
\Delta m = m_L-m_S =2 \mbox{Re}\, M_{12} 
\nnb
\]
where 
\be
M_{12}=\frac{G_F^2 M_W^2}{4\pi^2}\left\{
\lambda_c^2 \eta_1\frac{m_c^2}{M_W^2}+
\lambda_t^2 \eta_2\frac{m_t^2}{ M_W^2}+
2\lambda_c \lambda_t\frac{m_c^2}{ M_W^2}\ln\frac{m_t^2}{M_W^2}
\right\}
\left(\frac{\al_s^{''}(m_c^2)}{\al_s^{'''}(\mu^2)}\right)^{2/9}
\frac{\langle \bar K^0| \hat O| K^0 \rangle}{2m_K}
{}.
\ee

Here the standard notations were used: 
$\lambda_i= V_{id}V^*_{is}$, 
$i = u, c, t$ being the
elements of the Kobayashi-Maskawa matrix; $\eta_i$'s are some
coefficients due to
strong-interaction contributions, $\hat O$ is the four-quark operator
$\hat O = (\bar s_L\gamma_\mu d_L)^2$ with $\Delta S =2$.

Equation (1) was obtained with the help of the Wilson expansion at short
distances, so this contribution to the mass difference 
$\Delta m$ should be more
correctly called the contribution of short distances. Actually,
the existence of the other contributions to the mass difference 
was demonstrated by Wolfenstein \cite{3}. These 
new contributions are essentially  different from those
under discussion and come from long distances. We 
shall not dwell upon them and reserve the term ``mass difference'' and
notation $\Delta m$ for the short-distance contribution only.

The strong interactions enter eq.~(1) through Wilson's coefficient
$\eta_i$'s, which can be reliably calculated because of the property of
the asymptotic freedom of QCD, and through the matrix element of the
four-quark operator $\hat O$, being responsible for the $\Delta S=2$
transitions.  It
is this matrix element we are going to discuss.
It should be stressed that this problem is of gravely
nonperturbative nature and, thus, it is a hard nut to crack by
any of the existing techniques. 
However, because of the great importance of the numerical value
of the matrix element several attempts have been undertaken to evaluate
it.

Gaillard and Lee have estimated this matrix element by means 
of the vacuum saturation hypothesis with the result 
\[
\langle \bar K^0|\hat O|K^0 \rangle^{VS}=\frac{2}{3}f_K^2 m_K^2
{},
\]
$f_K$ being the $K$-meson decay constant. 
It is convenient to parameterized the exact matrix element with the help
of dimensionless parameter $B$ as follows:
\be
\langle \bar K^0|\hat O|K^0 \rangle=B\langle \bar K^0|\hat O|K^0 \rangle^{VS}
{}.
\ee

A lot of efforts has been made to bypass the vacuum saturation
hypothesis and to calculate the parameter $B$ independently, the
obtained results diverge from each other [4-12].

This work aims at giving an analysis of attempts to compute $B$ within the 
QCD sum rules approach.

First, Chetyrkin et al. \cite{12} applied the technique of the 
QCD sum rules to the
three-point correlator comprising the operator $\hat O$ and 
the interpolating currents
of the $K$-mesons with the result $\hat B = 1.2 \pm 0.1$. 
Second, after completion of that
work there appeared similar in its spirit the calculation by Pich and 
Rafael [10]. They used the two-point correlator of the 
four-quark operators and obtained $B = 0.38\pm  0.09$.

It is of interest to trace the cause of this marked difference 
since the sum rules 
method, as a rule, works with an accuracy of about $(20\div 30)\%$.

The work is organized as follows. In sect.~2 we briefly describe the calculation  
of $B$ by means of a three-point correlator and discuss the accuracy of the vacuum
saturation hypothesis employing there to estimate the vacuum expectation
values (v.e.v.) of four-quark operators. Section~3 is devoted to a comparison of
this calculation with the one performed by Pich and Rafael. Section 4 contains
our conclusions.

2. - The starting point of the work \cite{12} 
is the following representation:
\be
\langle \bar K^0|\hat O|K^0 \rangle=
\lim_{p^2\to m_K^2}\lim_{K^2\to m_K^2} 
(p^2- m_K^2)(K^2- m_K^2)p^\mu K^\nu 
\ee
\[
\phantom{MMMMMMM}
m_K^{-4}f_K^{-2}i^2 \int \dx \dy 
\langle
0|Tj^5_\mu(x)j^5_\nu(y)\hat O(0)|0 
\rangle
\exp[ipx-iKy]
{},
\]
which can be obtained with the reduction formulae. Thus, to
find the matrix element (3) one needs to compute the function
\be
T_{\mu\nu}(p,q) =i^2\int \dx \dy \exp[ipx-iqy]
\langle Tj^5_\mu(x)\hat O(y) j^5_\nu(0) \rangle=
\ee
\[
=p_\mu q_\nu T(p^2, (p-q)^2,q^2)+\mbox{other structures},
\]
where $j^5_\mu=\bar d\g_\mu\g_5 s$ is the interpolating field
of the $K^0$-meson:
\[
\langle 0|j^5_\mu(0)|K^0(p) \rangle=i p_\mu f_K,
\quad   f_K=1.17 f_\pi
{}.
\]

The function (4) can be reliably calculated at small $q$ and large
Euclidean 
$p^2  > 1~\GeV^2$ 
by means of the (somewhat modified) operator product expansion
technique \cite{13}. On the other hand, the matrix element (2) 
is connected with the
amplitude $T(-t,-t,0)\equiv T(t)$ by the dispersion relation
\be
T(t)=\int ds\frac{\rho(s)}{s+t}-\mbox{subtractions} =
f_K^2\frac{\langle \bar K^0|\hat O|K^0 \rangle}{(t+m_K^2)^2}+
\frac{A}{t+m_K^2}+\ldots
{},
\ee
where the one-pole contribution corresponds to transitions of
the $K^0$-meson to other (different from $K^0$) 
states and dots stand for the higher-state contributions.

Within the vacuum saturation approximation the function 
$T_{\mu\nu}(p,q)$ assumes the form
\be
T_{\mu\nu}^{VS} =\frac{8}{3}\Pi_{\mu\al}(p)\Pi_{\nu\al}(p-q),
\quad
\Pi_{\mu\al}(p)=i\int \dx \exp[ipx]
\langle Tj^5_\mu(x) \bar s_L(0)\g_\al d_L(0) \rangle
\ee
and the resulting value of $B$ proves to be $B^{VS}=1$. Thus, 
there remains to compute only the function
$\Delta_{\mu\nu}=T_{\mu\nu}-T_{\mu\nu}^{VS}$,
which is responsible for all the
departures from the vacuum saturation prediction for $B$. 

The computed result for $\Delta_{\mu\nu}$ is 
(only local operators with dimension $\le 6$ were taken into account) \cite{12}
\be
\Delta_{\mu\nu}(p,q)=p_\mu q_\nu \left\{-5 (pq) \frac{\langle \al_s
G^2 \rangle}{192\pi^3}-4\langle \bar d s \bar s d  \rangle 
-4\langle \bar d  d\bar s s   \rangle +
\right.
\ee
\[
\left.
+2\langle \bar s  s\bar s s   \rangle 
+2\langle \bar d  d\bar d d   \rangle 
+\frac{m_s \langle g\bar d G_{\mu\nu} \sigma_{\mu\nu}d
\rangle}{24\pi^2}
\right\}\frac{1}{p^2(p-q)^2}
+\mbox{other structures},
\]
where the designation $\langle \bar d s \bar s d  \rangle$ 
stands for $\langle \bar d_L\g^\al s_L \bar s_L \g^\al d_L \rangle$
and so on;
$G^2=G^a_{\mu\nu}G^a_{\mu\nu}$; $G_{\mu\nu}=G^a_{\mu\nu}t^a$;
$tr(t^a t^b)=(1/2)\delta^{ab}$;
$\sigma_{\mu\nu}=(i/2)[\g_\mu,\g_\nu]$.

On putting $q=0$ and using the method of finite energy sum
rules \cite{14} to connect the phenomenological (eq. (5)) and the theoretical
(eq. (7)) representations of the 
function $T(t)$, one gets 
\[
\frac{2}{3}  
f_K^4 m_K^2 (B - 1) = \int_0^{s_0} \rho^{th}(s) (s + m_K^2) ds 
= (24\pi^2)^{-1} m_s 
\langle g\bar d\sigma_{\mu\nu}G_{\mu\nu} d \rangle
\]
or 
\be
B-1=\frac{m_s \langle g\bar d \sigma_{\mu\nu}G_{\mu\nu} d
\rangle}{16\pi^2 f_K^4 m_K^2}
\ee

To estimate the v.e.v.'s of four-quark operators 
appearing in eq.~(7) the
hypothesis of vacuum saturation has been used, which leads to the vanishing of
each of these v.e.v.'s. Assuming the relation \cite{15}
\[
\langle g\bar d\sigma_{\mu\nu}G_{\mu\nu} d \rangle=m_0^2
\langle \bar d d \rangle \, ,\quad  m_0^2=(0.8\pm 0.4)~\GeV^2,
\]
one finds for the renormalization group invariant quantity
\footnote{
Having in mind a solid theoretical basis of our calculation we have just         
updated the prediction for $\hat B_K$ by using present values of                 
relevant parameters. The quantity $B-1$ in eq.~(8) is expressed through          
well known parameters that did not change much during last years.                
The bulk of the change of the parameter $\hat B$                                 
is due to the normalization (the change in the factor                            
$\al_s(1.2~{\rm GeV}^2)^{-2/9}$ in eq.~(9)).                                     
The updated version of eq.~(9) with $\al_s(1.2~{\rm GeV}^2)=0.69$                
reads now:                                                                       
\[                                                                               
\hat B = 1.0\pm 0.1                                                              
\]                  
}
\be
\hat B = B(\mu) (\al_s(\mu))^{-2/9}=1.2\pm 0.1
{}.
\ee

This estimate is in good agreement with the hypothesis of vacuum saturation;
within this approach this means the absence of operators giving appreciable
contributions to $\Delta_{\mu\nu}$.

In principle there may exist three kinds of extra corrections to  
$\Delta_{\mu\nu}$:

i) corrections of higher order in $\al_s$;

ii) corrections due to bilocal operator;

iii) corrections due to local operators.

Let us consider them in turn. The first contribution is suppressed by the
factor $\al_s/\pi \sim 0.1$ and can hardly change the result (9) 
to a large extent. The
second one is suppressed by an extra power of $p^{-2}$ 
and does not contribute at all
within the finite energy sum rules approach. (This statement
is valid as long as
one neglects, as we always do, a weak logarithmic dependence
on $p^2$ induced by
anomalous dimensions of the operators involved.) The third type of terms
have been taken into account in eq.~(7) in the leading order with a
small total effect on
$(B - 1)$. Thus, all the contributions are under control within
the approach and prove to be small.

However, there exists another source of uncertainty -- the use of 
the procedure of vacuum saturation to evaluate the v.e.v.'s of 
four-quark operators,
which might (as simple estimates show) violate drastically eq.~(9). Let
us discuss the issue in some detail.   

To begin with, it is easy to show that the peculiar combination
of four-quark 
operators entering eq.~(7) transforms as a member of a 27-plet
with respect to
the (flavour) $SU^f(3)$ group. 
This means that the corresponding v.e.v's
might only appear in the second order of the expansion in 
the strange-quark mass. It
can be checked that this suppression is still operative if one takes into account
next-to-leading corrections to the coefficient functions 
of the operators under
consideration. However, it is interesting to get a direct estimate 
of the accuracy of
the vacuum saturation hypothesis. To this end it is convenient to employ the
QCD sum rules method in configuration space ($x$-space). 
It will be seen below
that such $x$-space sum rules provide the unique possibility of 
direct estimation of
the v.e.v.'s of four-quark operators.

Indeed, let us consider the correlator
\[
\langle T j^\mu(x) j_\mu(0)\rangle =\Pi(-x^2)
{},
\]
where $ j^\mu= \bar u \g^\mu d $ is the interpolating current
of the $\rho$-meson. 
At $x^2\to 0$ the
following $x$-space operator expansion holds:
\be
\Pi(-x^2)=\frac{6}{\pi^4 x^6}
+\frac{\langle \al_s G^2 \rangle}{16\pi^3 x^2}+\langle j^\mu
j_\mu \rangle
+\frac{\al_s}{4\pi}\left[6\left(L+2\g_E+\frac{1}{2}\right)
\langle \bar u \g_5\g^\lambda t^a d  \bar d \g_5\g^\lambda t^a
u   \rangle + 
\right.
\ee
\[
\left.
+\left(\frac23 L+\frac43 \g_E -\frac{13}{9}\right)
\langle \bar \psi \g^\lambda t^a \psi
(  \bar u \g^\lambda t^a u  +
 \bar d \g^\lambda t^a d) \rangle 
\right]
+o(1)
\]
where $L =\ln (-\mu^2 x^2/4)$, $\g_E = 0.577\ldots$, 
and $\mu$ is the normalization point of the
$\msbar$-scheme. Expansion (10) can be obtained from the
corresponding one in $p$-space. 
Note that the values of the constants added to $L$
in the right-hand side of
eq.~(10) are fixed by the recipe for renormalization of the
operator $j^\mu j_\mu $. Note also
that expansion (10) remains valid after substitutions $j^\mu\to j^\mu_5$, 
$\g_5\g_\lambda\to \g_\lambda$,  
which
correspond to considering the axial current interpolating the $A_1$-meson.

On the other hand the following Kallen-Lehmann representation takes place:
\be
\Pi(-x^2)=\int_{4m_\pi^2}^\infty \rho(s)
\Delta^c(-x^2,s)ds, \quad  \Delta^c(x^2,s)
=\frac{1}{4\pi^2 x^2}\sqrt{x^2 s} \, K_1(\sqrt{x^2 s}),
\ee
where the spectral density can be roughly approximated as follows:
\[
\rho(s) = F\delta(s - m^2) + a\theta(s - s_0){},
\]
the parameters $F$, $m$, $s_0$, $a$ being known 
from the analysis of the same correlator
in $p$-space \cite{16}.   

The main advantage of relations (10) and (11) over similar sum rules
in $p$-space
comes from the fact that the contribution of the operator
$j^\mu j_\mu$ on the right-hand
side of eq.~(10) is parametrically increased due to the
contact term pictured in fig.~1 
and corresponding to a disconnected diagram. This allows one to use
eqs.~(10) and (11) for determining the v.e.v. 
$\langle j^\mu j_\mu \rangle$  directly.

\parbox{\textwidth}{
\begin{center} 
\begin{picture}(350,80)(0,0)
\Line(50,0)(30,60)
\Line(50,0)(70,60)
\put(50,0){\circle*{4}}
\Text(52,-10)[c]{x }

\Line(120,0)(100,60)
\Line(120,0)(140,60)
\put(120,0){\circle*{4}}
\Text(122,-10)[c]{0}

\put(150,20){x$\,\,-\!\!-\!\!\!\!\longrightarrow\, 0$}
\Line(250,0)(220,60)
\Line(250,0)(260,60)
\Line(250,0)(290,50)
\Line(250,0)(310,30)
\put(250,0){\circle*{4}}
\Text(250,-10)[c]{0}
\end{picture}
\end{center}
\vspace{.8cm}

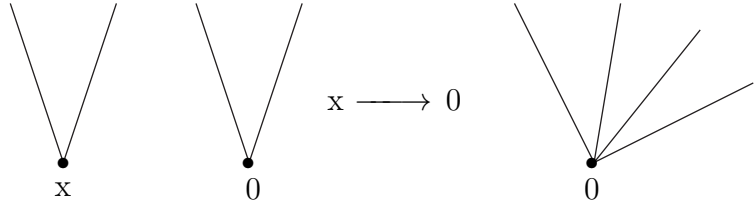
Fig.~1 - The origin of the contact terms in the 
$x$-space operator product expansion.

\label{fig1}      
}
\vspace{.4cm}

An analysis of sum rules in $x$-space leads to the relations
(see fig.~2 and 3; we took $\mu=1~\GeV$, $\al_s(1 \GeV)=0.3$,
$\langle \bar qq \rangle=(-0.23~\GeV)^3$)
\[
\langle \bar u\g^\mu d \bar d \g^\mu u \rangle/\langle \bar qq
\rangle^2
=-\frac{1}{3}(0.90\pm 0.15)\, ,
\]
\[
\langle \bar u\g^\mu \g_5d \bar d \g^\mu \g_5 u \rangle/\langle \bar qq
\rangle^2
=\frac{1}{3}(0.84\pm 0.20)\, ,
\]

\parbox{\textwidth}{
\begin{center} 
\begin{picture}(350,100)(0,0)
\Line(30,0)(30,5)

\Text(30,-10)[c]{ 2.0 }
\Text(100,-10)[c]{ 2.5 }
\Text(170,-10)[c]{ 3.0 }
\Text(240,-10)[c]{ 3.5 }

\Line(0,0)(260,0)
\Line(0,0)(0,75)

\Text(30,-10)[c]{ 2.0 }
\Text(100,-10)[c]{ 2.5 }
\Text(170,-10)[c]{ 3.0 }
\Text(240,-10)[c]{ 3.5 }

\multiput(0,0)(0,25){4}{\Line(0,0)(5,0)}

\Text(-15,0)[c]{0.6}
\Text(-15,25)[c]{0.8}
\Text(-15,50)[c]{1.0}
\Text(-15,75)[c]{1.2}

\Curve{(15,75)(100,40)(170,23)(240,21)}
\end{picture}
\end{center}

\vspace{.8cm}

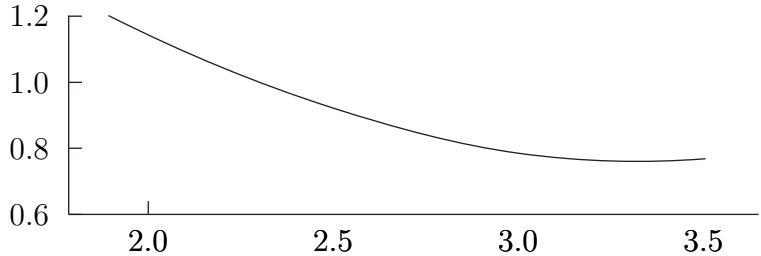
Fig.~2 -  The curve showing the dependence of the  numerical value of 
the v.e.v.  \\ 
$
\displaystyle
\langle \bar u \g^\mu  \bar d d \g^\mu  u \rangle 
$ 
on the variable $|x| = \sqrt{-x^2}$.

\label{fig2}  }
\vspace{.8cm}


\parbox{\textwidth} {
\begin{center} 
\begin{picture}(350,100)(0,0)

\multiput(0,0)(70,0){4}{\Line(30,0)(30,5)}

\Line(30,0)(30,5)

\Text(30,-10)[c]{ 2.0 }
\Text(100,-10)[c]{ 2.5 }
\Text(170,-10)[c]{ 3.0 }
\Text(240,-10)[c]{ 3.5 }

\Line(0,0)(260,0)

\Line(0,0)(0,80)
\Line(0,15)(5,15)
\Text(-15,15)[c]{0.6}

\multiput(0,0)(0,25){3}{\Line(0,15)(5,15)}

\Text(-15,15)[c]{0.6}
\Text(-15,40)[c]{0.8}
\Text(-15,65)[c]{1.0}

\Curve{(20,76)(30,70)(100,40)(240,21)}
\end{picture}
\end{center}

\vspace{.8cm}

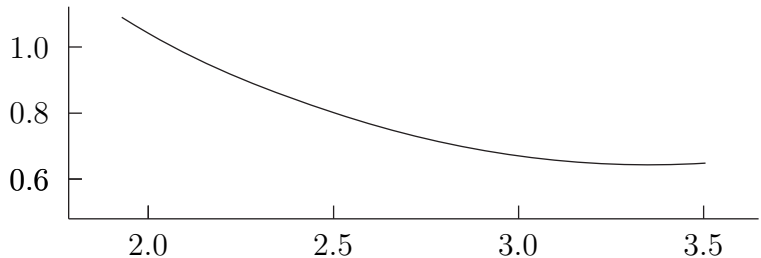
Fig.~3 - The curve showing the dependence of the  numerical value of 
the v.e.v.  \\
$
\langle \bar u\g^\mu \g_5d \bar d \g^\mu \g_5 u \rangle
$
\label{fig3}      
on the variable $|x| = \sqrt{-x^2}$
{}.
}
\vspace{.8cm}\\
which are in agreement with the result of the vacuum
saturation within an
accuracy 30\%. The relative discrepancy between v.e.v.'s of
four-quark operators of $VV$ and $AA$ types is about 10\%; this
fact indicates in favor of a small contribution of four-quark
operators entering the right-hand side of eq.~(7) to the value
of $B$.

3. - Let us discuss another calculation of $B$ 
\cite{10} employing the finite energy sum
rules for the two-point correlator          
\[
P(Q^2)=i\int \dx \exp[iqx] \langle 0| T \hat O(x) \hat O(0) |0\rangle
\]
together with the method of effective chiral Lagrangians, the latter being used
to fix the functional dependence of the corresponding spectral
density on energy.
The obtained value of $B$ differs noticeably from both result (9) 
and the prediction
of the vacuum saturation hypothesis. It should be noted that this calculation
involved a large-size correction due to the operators $m^2$,
$m^4$, $m\bar qq$. This fact
forced the authors to use a rather large ``duality interval'' 
$S_0 = 8 \GeV^2$ in handling
their sum rules. On the other hand, within the approach of 
paper \cite{12} all these
contributions are exactly summed up in the factorized term. Note also that the
use of the chiral perturbation theory assumes the independence of the form
factor from the energy of the $K^0 -\!\!\!\!- \bar K^0$ pair, 
which can hardly be a reasonable
approximation at energies as  large as  $\sqrt{S_0}=(2.5\div 3)\GeV$. The
technicalities aside, we feel that the result of Pich and Rafael may 
well be somewhat underestimated.

4. - To conclude, we have used the $x$-space sum rules to get a direct
estimation of the accuracy of the vacuum saturation technique
for operators $\bar u\g^\mu d \bar d \g^\mu u$
and $\bar u\g^\mu \g_5d \bar d \g^\mu \g_5 u $. 
Our result is that the v.e.v.'s of these operators
are equal to each other with an accuracy of about 10\%. 
This observation gives an
additional support of the result of the calculation of the parameter 
$B$ which was discussed in sect.~2.

\end{document}